# High-speed Fourier-transform infrared spectroscopy with phase-controlled delay line

Kazuki Hashimoto[1], Venkata Ramaiah Badarla[1] and Takuro Ideguchi[*1,2]

[1] Institute for Photon Science and Technology, The University of Tokyo, Tokyo 113-0033, Japan

[2] PRESTO, Japan Science and Technology Agency, Saitama 332-0012, Japan

*ideguchi@ipst.s.u-tokyo.ac.jp

**Abstract**

**Fourier-transform infrared spectroscopy (FTIR) is the golden standard of mid-infrared (MIR) molecular spectroscopic analysis through optically-encoded vibrational signatures. Michelson-type FTIR and MIR dual-comb spectrometers allow us to simultaneously investigate multiple molecular species via the broadband and high-resolution spectroscopic capabilities. However, these are not applicable to high-speed measurements due to the low temporal resolution which is fundamentally limited by the signal-to-noise ratio (SNR). In this study, we develop a high-speed FTIR spectroscopy technique called phase-controlled Fourier-transform infrared spectroscopy (PC-FTIR) that has the capability to measure MIR absorption spectra at a rate of above 10 kHz. PC-FTIR demonstrates the high scan rate with a high SNR for various spectral bandwidths by arbitrarily adjusting the instrumental spectral resolution. As a proof of principle demonstration, we measure high-speed mixing dynamics of two liquids at a rate of 24 kHz. We also measure MIR spectra of gas-phase molecules with higher spectral resolution at a rate of 12 kHz. This high-speed MIR spectrometer could be especially useful for measuring non-repetitive fast phenomena and acquiring a large amount spectral data within a short time.**

## 1. Introduction

Fourier-transform infrared spectroscopy (FTIR) has been widely used in a variety of fields as a simple and robust tool for label-free molecular analysis. Michelson-type FTIR spectrometers with a thermal source can simultaneously interrogate multiple molecular species via an ultra-broadband mid-infrared (MIR) spectrum covering the entire fundamental molecular vibrational region (400-4,000cm$^{-1}$) [1]. Recent advancement in nonlinear wavelength-conversion techniques with ultrashort pulsed lasers (e.g. difference-frequency generation (DFG) or optical parametric oscillation (OPO)) has provided new opportunities to exploit the broadband and/or highly-tunable bright MIR sources for FTIR spectrometers [2-5]. For example, dual-comb spectroscopy (DCS) with two mutually coherent frequency combs based on the MIR laser sources significantly improved the frequency accuracy and also spectral resolution down to ~100 MHz (~0.003 cm$^{-1}$) [6-10], benefiting precision molecular analysis. These broadband and high-resolution FTIR spectrometers are generally applied only for measuring static samples because they require data averaging with a long measurement time to achieve sufficiently high signal-to-noise ratio (SNR). Therefore, they are not suitable for measuring non-repetitive fast phenomena such as flowing cells/particles [11,12] or gaseous combustion [13,14]. To achieve a further higher temporal resolution, we have to improve the SNR by adjusting the spectroscopic parameters, which are governed by the trade-off relation [15] given as: $SNR \propto T^{\frac{1}{2}} D^{\frac{1}{2}} \Delta\nu^{-1} \nu_{\text{res}}$, where $T$, $D$, $\Delta\nu$ and $\nu_{\text{res}}$ denote measurement time, duty cycle, spectral bandwidth, and spectral resolution, respectively.

To obtain a high SNR within a short measurement time, reducing spectral resolution and/or bandwidth is required according to the above-mentioned trade-off relation. One approach for this is DCS with MIR comb sources with higher repetition rates of ~10 GHz - ~100 GHz (which corresponds to instrumental spectral resolutions of ~0.3 cm$^{-1}$ - ~3 cm$^{-1}$) such as quantum cascade lasers [16], interband-cascade lasers [17], micro-resonators [18] and electro-optic modulators [19]. Demonstrations of MIR-DCS with those frequency comb sources have shown the scan rate of up to about 1 MHz [20-22] and used for measuring fast phenomena such as protein reactions [23], liquid flow [24] and high-temperature gaseous reaction [25] with their high temporal resolution. Although these DCS spectrometers are promising approaches, they require exotic comb sources with limitations imposed by the low pulse energy, limited spectral tunability, and/or low spectral flatness. Another approach is DCS with low repetition rate (~100 MHz) MIR combs generated by the nonlinear wavelength conversion techniques (DFG or OPO combs). By truncating an interferogram within a short time window can effectively reduce the spectral resolution, making the SNR and temporal resolution higher by avoiding data averaging [26-28]. However, the interferogram truncation causes reduction of measurement duty cycle: for example, 99-99.9% of interferogram data points are discarded to achieve 10-100 GHz spectral resolution with a 100-MHz DCS system. To benefit the high pulse energy, high wavelength tunability and/or high spectral flatness of the MIR laser sources generated by the nonlinear wavelength conversion techniques, an efficient FTS technique with a higher duty cycle is demanded.

In this study, we develop phase-controlled FTIR (PC-FTIR), with which we can measure MIR spectra at a scan rate of 10s kHz with a wavelength-conversion-based MIR laser source running at a repetition rate of 10s-100s MHz. We use a wavelength-tunable free-running MIR femtosecond OPO as a light source in a high-speed phase-controlled Fourier transform spectroscopy (PC-FTS) system, which has been recently proposed and demonstrated in the near-infrared region [29]. Since PC-FTIR is an autocorrelation-type FTIR spectrometer, we can arbitrarily set the spectral resolution (e.g. ~0.3 cm$^{-1}$, ~3 cm$^{-1}$) independent of the laser repetition rate without sacrificing the duty cycle. We also emphasize that the autocorrelation configuration allows us to use a single free-running MIR source without the need for any sophisticated laser stabilization techniques. As a proof of principle demonstration, we measure mixing dynamics of organic liquids with the broadband MIR spectra at a spectral acquisition rate of 24 kHz. In addition, we measure gas-phase molecular absorption with a higher spectral resolution and relatively narrower MIR spectra at a spectral acquisition rate of 12 kHz by adjusting the spectral resolution and bandwidth to keep the high SNR (measurement time).

## 2. Methods
### 2.1 PC-FTIR
A schematic of PC-FTIR is shown in **Figure 1**. The system consists of a home-made femtosecond MIR-OPO and a phase-controlled Fourier-transform spectrometer. A tunable (690-1020 nm) Kerr lens mode-locked Ti:Sapphire oscillator (Maitai, Spectra Physics) working at a repetition rate of 80 MHz and with a pulse duration of <100 fs (Maitai, Spectra Physics) is used as a pump source for the OPO. MgO:PPLN (HC Photonics) with a fanout grating period of 19 - 23 μm is used as a non-linear medium in a singly resonant Z-cavity configuration. The generated idler pulses are separated from the pump pulses with a Ge long-pass filter. The spectral bandwidth and broad tunability of the idler pulses are around 200 cm$^{-1}$ and 2000-4500 cm$^{-1}$, respectively, as shown in the inset of **Figure 1**. The MIR idler pulses are coupled into an InF$_3$ single mode fiber for the spatial mode cleaning and collimated with a lens. The broadband MIR pulses are steered into a Michelson interferometer with a phase-controlled delay line (PCDL) [30]. The PCDL consists of a grating, a curved mirror and a scanner placed in the 4-f configuration. While the focal length of the curved mirror is fixed at 150 mm, the groove density of the grating is switchable from 180 lines/mm to 40 lines/mm. The 180-lines/mm and 40-lines/mm gratings are utilized for gas-phase (higher resolution) and condensed-phase (lower resolution) measurements, respectively. A 12-kHz resonant scanning mirror (CRS12kHz, Cambridge Technology) is used as a rapid scanner. The PCDL generates group delay of the pulses in the scan arm by adding linear spectral phase with a scanned angle of the scanner placed in the Fourier plane. The amount of delay is continuously scanned at a rate of 24 kHz via the round-trip motion of the 12-kHz resonant scanner. The maximum group delay, which depends on the system configuration, determines the spectral resolution. In the reference arm, a neutral density filter is inserted to adjust the power ratio between the two arms. The MIR pulses from the scan and reference arms are recombined at the output port of the beamsplitter in the interferometer, interacted with a sample, and detected with a N$_2$-cooled MCT detector (KMPV8-0.5-J1/DC50, Kolmar technologies) after passing through a polarizer. The detected signal is low-pass filtered, amplified, and digitized (ATS9440, AlazarTech). The recorded temporal waveform is segmented into individual interferograms. Since the PCDL generates nonlinear group delay in time, a post correction method for linearization [29, 31] is applied. Finally, Fourier-transforming the corrected interferograms yields MIR spectra.

For measuring mixing dynamics of condensed-phase molecules, we use two liquid samples: phenylacetylene and toluene. The phenylacetylene and toluene are mixed in a homemade cuvette consisting of a pair of 3-mm-thick KBr windows with a 200-μm-thick Teflon spacer. During the measurement, phenylacetylene is injected into the cuvette filled with toluene. The MIR pulses are focused onto the sample with a f=30 mm CaF$_2$ lens, and the transmitted pulses are collected with another lens with the same focal length. For gas-phase MIR spectroscopy, we use 20-Torr $^{12}C^{16}O$ molecules filled in a gas cell with a length of 50 mm. The collimated MIR beam passes through the cell in a round-trip geometry.

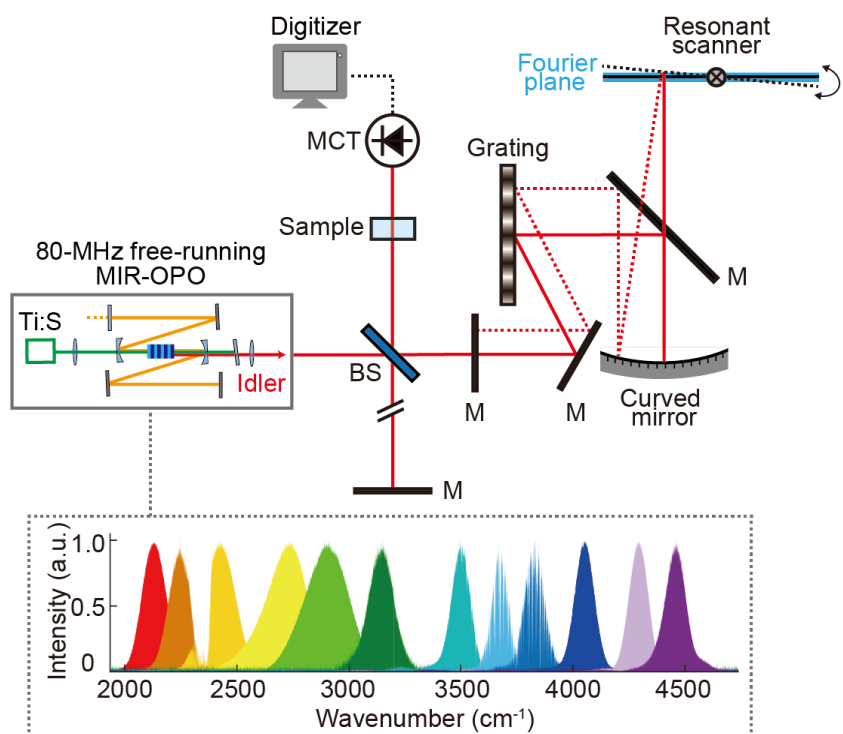

**Figure 1** Schematic of PC-FTIR. MIR-OPO: mid-IR optical parametric oscillator, BS: beamsplitter, M: mirror, MCT: HgCdTe detector. The inset shows the idler spectra of the MIR-OPO measured by a home-made FTIR spectrometer.

## 3. Results

### 3.1 High-speed MIR spectroscopy of mixing dynamics of liquids

We measure mixing dynamics of liquid phenylacetylene and toluene by PC-FTIR in order to show the applicability of the system for high-speed condensed-phase MIR spectroscopy. **Figure 2a** shows sequentially obtained interferograms. The interferograms are acquired at a rate of 24 kHz as shown in the inset of **Figure 2a**. The peak intensities of the interferograms vary depending on the concentration of the liquids at the focal volume. The rapidly varying spikes around 39 ms in the interferograms would originate from contaminated bubbles. **Figure 2b** shows FFT spectra of the double-sided interferograms. The MIR spectral range is adjusted at 2050 - 2250 cm$^{-1}$, and the triangular-apodized spectral resolution is 5.1 cm$^{-1}$. The absorption peaks appeared at 2112 cm$^{-1}$ and 2165 cm$^{-1}$ originate from C≡C stretching of phenylacetylene and C=C stretching overtone band of toluene, respectively [32, 33]. To avoid the detector nonlinearity, the average power illuminated on the detector is restricted up to around 1.5 μW, which can be improved to more than 10 μW by using a higher dynamic range detector. Under this measurement condition, SNR of a single-shot non-averaged spectrum is 36 at the peak of the spectrum, where the detector noise is dominant. **Figure 2c** shows 6-averaged MIR spectra. The gray lines plotted together with the spectra represent the MIR spectra measured without the sample. The two absorption peaks vary in time depending on the concentration of the two species. **Figure 2d** describes temporal absorbance change of the two MIR bands plotted at a temporal resolution of 417 µs. The absorbance at the two wavenumbers complementarily changes to each other. Note that the absorbance changes at 2165 cm$^{-1}$ are much smaller than those at 2112 cm$^{-1}$ because the absorption coefficient of phenylacetylene at 2112 cm$^{-1}$ is stronger than that of the overtone band of toluene at 2165 cm$^{-1}$.

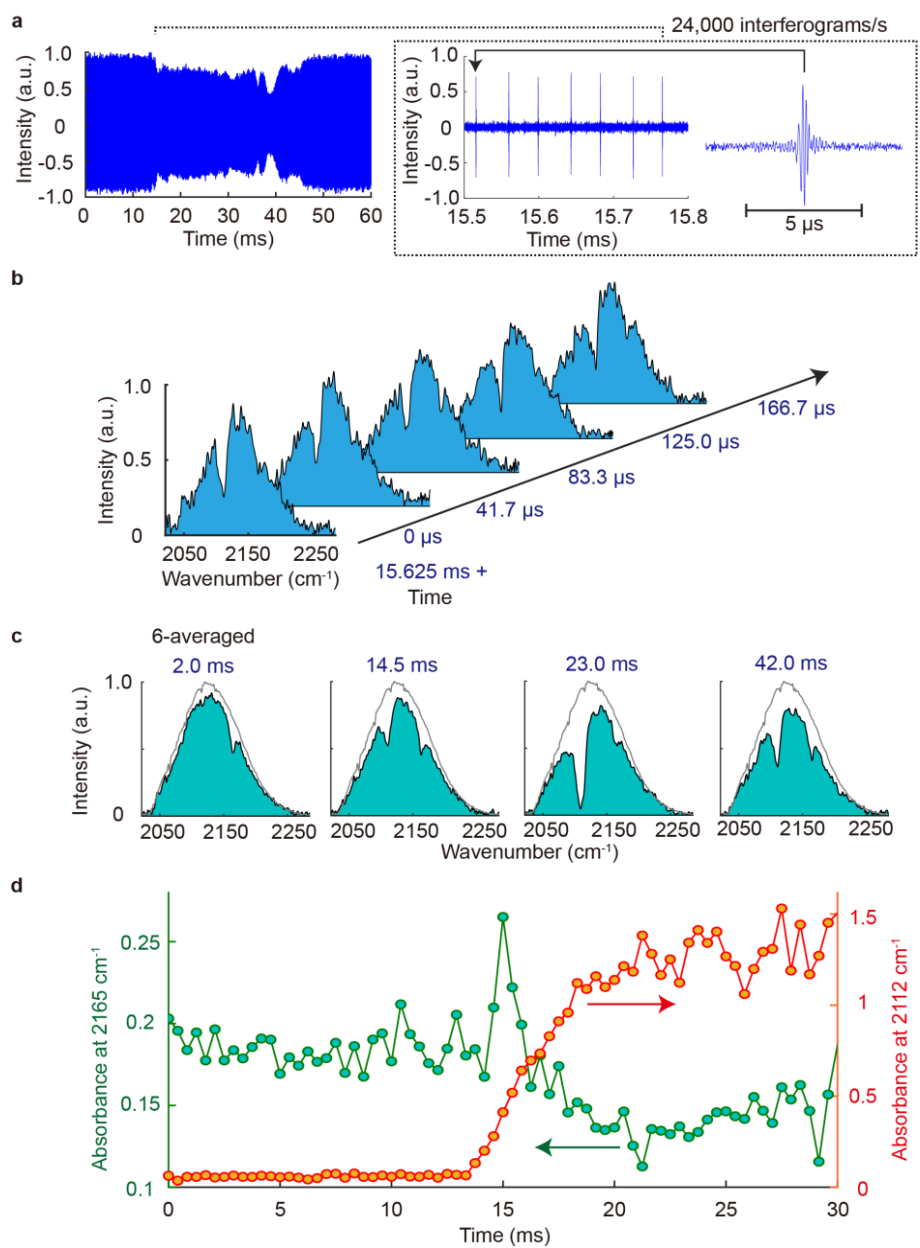

**Figure 2** High-speed MIR spectroscopy of mixing dynamics of two liquid molecules by PC-FTIR. a) Sequentially obtained MIR interferograms over 60 ms at a scan rate of 24 kHz. The inset shows an enlarged view of the interferograms. b) Continuous MIR spectra obtained at a rate of 24 kHz. The triangular-apodized spectral resolution is set to 5.1 cm$^{-1}$ (double-sided). The absorption peaks at 2112 cm$^{-1}$ and 2165 cm$^{-1}$ originate from C≡C stretching of phenylacetylene and C=C stretching overtone band of toluene, respectively. c) Six-averaged MIR spectra. The gray lines represent the averaged MIR spectra measured without the sample. d) Temporal absorbance change of the two

MIR bands during the mixing process (Orange: 2112 cm$^{-1}$ Green: 2165 cm$^{-1}$) obtained at a temporal resolution of 417 μs.

## 3.2 High-speed gas-phase MIR spectroscopy

We also measure gas-phase $^{12}C^{16}O$ molecules in a cell and $^{12}C^{16}O_2$ molecules in the ambient air with the PC-FTIR spectrometer. **Figure 3a** shows continuously measured MIR spectra at a rate of 24 kHz with an unapodized resolution of 2.9 cm$^{-1}$. The large absorption bands of $^{12}C^{16}O_2$ molecules in the ambient air appear at 2300-2400 cm$^{-1}$. By tuning the idler wavelength of the OPO, we can obtain MIR spectra at different center wavenumbers as shown in **Figure 3b**. The absorption bands of $^{12}C^{16}O$ molecules appear at the lower wavenumber region. Note that the spectral bandwidth of these spectra is limited by the idler's spectral bandwidth, while the instrumental bandwidth of the spectrometer is ~270 cm$^{-1}$. In this demonstration, the absorption bands of two gaseous molecules are observed, but the spectral resolution of 2.9 cm$^{-1}$ is not sufficient for resolving the sharp absorption lines of the gaseous molecules.

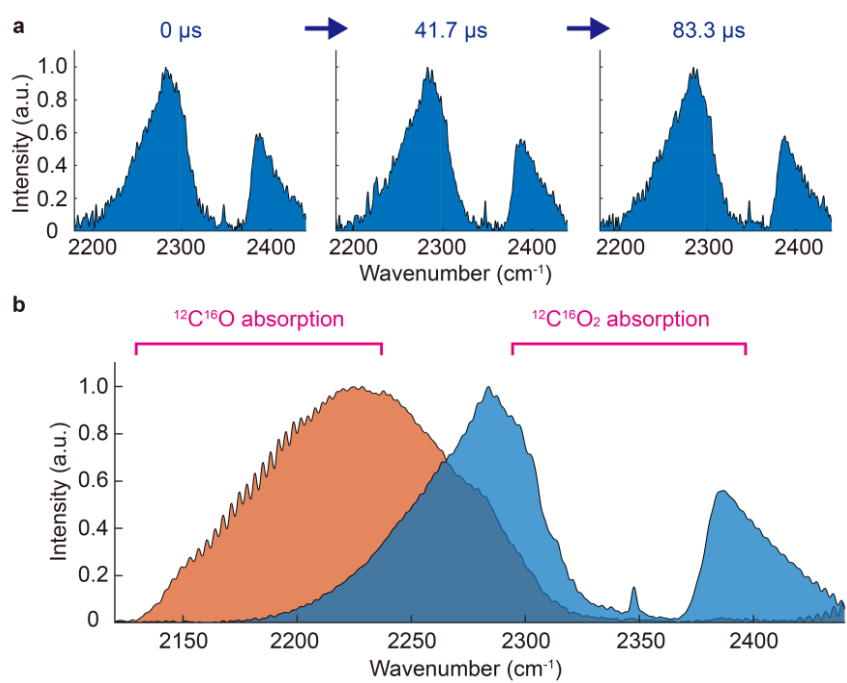

**Figure 3** Broadband MIR gas-phase spectra measured by PC-FTIR. a) Sequentially obtained MIR spectra at a rate of 24 kHz with the unapodized spectral resolution of 2.9 cm$^{-1}$. The large absorption bands originate from $^{12}C^{16}O_2$ molecules in the ambient air. b) Broadband 200-averaged spectra covering different wavenumber regions obtained by tuning the idler wavelength of the OPO. The absorption lines of $^{12}C^{16}O$ molecules in a gas cell also appear. The spectral bandwidth is limited by the idler's bandwidth of 200 cm$^{-1}$, while the system can accept the bandwidth of ~270 cm$^{-1}$.

To clearly resolve the sharp absorption lines of the gaseous molecules, we set the spectral resolution higher by narrowing the spectral bandwidth. Note that we keep the scan rate at the same value as before. A grating with a larger number of groove density and an optical band-pass filter are implemented to adjust the spectral resolution and bandwidth, respectively. **Figure 4a** shows interferograms of gas-phase $^{12}C^{16}O$ molecules obtained at the higher-resolution configuration. As shown in the left panel, interferograms are sequentially measured at a rate of 24 kHz. The 200-averaged interferogram is shown in the right panel. The molecular vibrational free induction decay of $^{12}C^{16}O$ molecules is clearly observed as shown in the inset. The maximum group delay of 120 ps is scanned within 27 μs. The duty cycle is set to 0.65 (27 μs / 41.7 μs) because visibility of the interferogram decreases at large scan angles possibly due to the off-axis aberration in the PCDL. Satellite small peaks around the interferograms caused by the multiple reflections from the beamsplitter in the interferometer are numerically suppressed. Sequentially obtained spectra with a resolution of 0.29 cm$^{-1}$ at a rate of 12 kHz are shown in the left panels in **Figure 4b**. Each spectrum is obtained by Fourier-transforming a two-averaged single-sided interferogram with the nonlinear delay correction. The sharp absorption lines of $^{12}C^{16}O$ molecules are clearly observed in the spectra. The 200-averaged spectrum shows clearer absorption lines as shown in the right panel of **Figure 4b**. We note that state-of-the-art rapid-scan FTIR spectrometers have demonstrated a scan rate of ~80 kHz at the spectral resolution of ~10 cm$^{-1}$ [34], while it decreases down to less than 10 Hz at a high resolution of around 0.3 cm$^{-1}$ [35]. Therefore, the scan rate of the PC-FTIR spectrometer is higher than the previous methods by a factor of more than 1000.

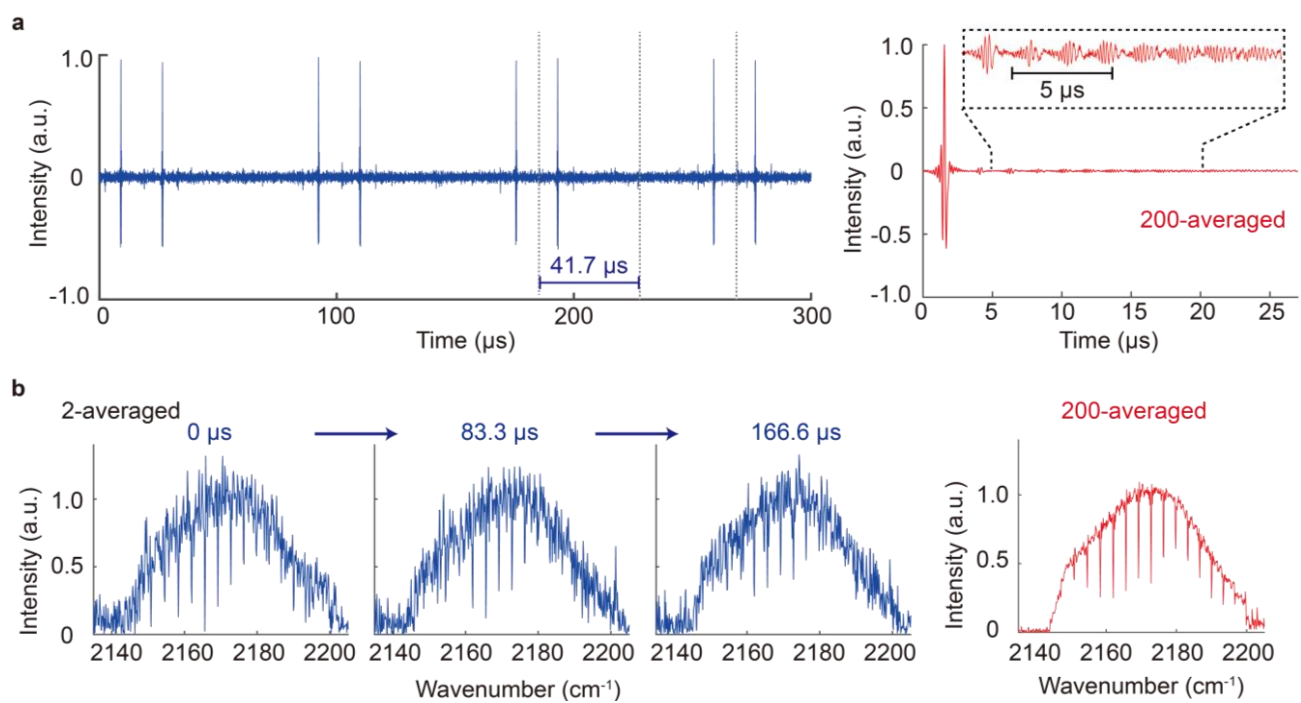

**Figure 4** High-resolution MIR spectra of gas-phase molecules measured at a rate of >10 kHz. a) Interferograms measured at the high-resolution configuration. Blue: continuous interferograms obtained at a rate of 24 kHz. Red: A 200-averaged interferogram. The inset shows an enlarged view of the interferogram. b) FFT spectra of the interferograms. Blue: $^{12}C^{16}O$ spectra measured with a resolution of 0.29 cm$^{-1}$ at a scan rate of 12 kHz. Red: A 200-averaged spectrum.

## 4. Conclusion

We demonstrated high-speed condensed- and gas-phase PC-FTIR spectroscopy at a rate of >10 kHz by arbitrarily adjusting the spectral resolution and bandwidth. Our system can be improved in various directions. The current SNR of the spectra is limited by the low dynamic range of our detector. If the SNR becomes higher with a higher-dynamic-range detector, the scan rate can be increased by using a polygonal scanner, whose scan rate can be up to 50 kHz [36]. We can also use other types of broadband MIR sources such as DFG-based broadband MIR lasers. Moreover, since we use an ultrashort pulsed laser as a light source, the MIR spectroscopy by PC-FTIR can be simultaneously implemented with another nonlinear spectroscopic modality such as coherent Raman scattering spectroscopy [37]. Furthermore, contrary to using the pulsed lasers, we can use temporally incoherent light sources such as sun light, thermal lamp or quantum cascade superluminescent emitter [38] as a MIR light source because PC-FTIR works as a passive spectrometer that is able to measure an external light source. It could be used as a field deployable compact spectrometer. Finally, the high-speed FTIR spectrometer will become a powerful tool for various analytical applications such as liquid/particle flow measurements [11,12], gaseous combustion measurements [13,14], large-area remote sensing [39, 40] and photoacoustic sensing [41].

**Acknowledgements**

We thank Yu Nagashima for use of his equipment. This work was financially supported by JSPS KAKENHI (20H00125), and JST PRESTO (JPMJPR17G2).